\documentclass[pdflatex,sn-mathphys-num]{sn-jnl}% Math and Physical Sciences Numbered Reference Style
\usepackage{amssymb,amsmath,amsfonts}
\usepackage{mathtools,cuted}
\usepackage{dsfont}
\usepackage[utf8]{inputenc}
\usepackage{flushend}
\usepackage[T1]{fontenc}
\usepackage{graphicx,psfrag,color}
\usepackage{dcolumn}
\usepackage{bm}
\usepackage{mathbbol}
\usepackage{braket}

\theoremstyle{thmstyleone}%
\theoremstyle{thmstyletwo}%

\theoremstyle{thmstylethree}%

\begin{document}

\title[Barenco gate implementation using driven two- and three-qubit spin chains]{Barenco gate implementation using driven two- and three-qubit spin chains}

\author[1]{\fnm{Rafael} \sur{Vieira}}\email{rafael.vieira1989@edu.udesc.br}
\equalcont{These authors contributed equally to this work.}

\author*[1]{\fnm{Edgard P. M.} \sur{Amorim}}\email{edgard.amorim@udesc.br}
\equalcont{These authors contributed equally to this work.}

\affil*[1]{\orgdiv{Departamento de F\'isica}, \orgname{Universidade do Estado de Santa Catarina}, \orgaddress{\postcode{89219-710}, \city{Joinville}, \state{SC}, \country{Brazil}}}

\abstract{We propose a protocol for implementing Barenco-type multi-qubit controlled gates using short driven spin chains. Starting from an Ising interaction with a transverse drive on the last spin, we construct an effective two-qubit Hamiltonian whose time evolution implements the Barenco gate $V_2(\varphi,\omega,\phi)$ and, in particular, a CNOT gate. We then embed this construction into a three-qubit $XXZ$ chain to realize the three-qubit Barenco gate $V_3(\varphi,\omega,\phi)$, which includes the Toffoli gate as a special case. The derivation is fully analytical: we perform a sequence of unitary transformations, identify decoupled subspaces, and apply a rotating-wave approximation to obtain simple effective Hamiltonians. We derive explicit conditions on the coupling strengths and driving parameters, provide closed-form expressions for the time-evolution operators in each relevant subspace, and characterize the quality of the implementation using the operator fidelity. Numerical simulations show that the protocol achieves high fidelities over broad parameter ranges, demonstrating its robustness and suitability for quantum information processing in spin-chain platforms.}

\keywords{Quantum gate, Barenco gate, Toffoli gate, Spin chain}


\maketitle

%%%%%%%%%%%%%%%%%%%%%%%%%%%%%%%%%%%%%%%%%%%%%%%%%%%%%%%%%%%%
\section{Introduction}\label{sec:introduction}

Quantum technologies rely critically on the ability to implement high-fidelity few-qubit gates in a controllable and scalable manner~\cite{Ben00}. In the standard circuit model, universality can be achieved with arbitrary single-qubit rotations and a single nontrivial two-qubit gate, such as the controlled-NOT (CNOT) gate~\cite{Nie00}. However, many algorithms and error-correcting codes benefit from multi-qubit controlled operations, such as the three-qubit Toffoli gate, that act nontrivially only on a small subspace of the full Hilbert space.

Barenco~\cite{Bar95} introduced a particularly convenient family of multi-qubit controlled gates $V_N(\varphi,\omega,\phi)$ in which $N-1$ control qubits conditionally drive a single target qubit through a general single-qubit rotation. For suitable choices of the real parameters $(\varphi,\omega,\phi)$, this family reproduces standard gates such as CNOT and Toffoli~\cite{Bar95,Bar95a,Tof80,Ras20}. Because of their tunability and simple algebraic structure, Barenco gates are natural building blocks for quantum algorithms and quantum simulators~\cite{Lee04,Shi18}.

From a physical perspective, both short and long spin chains constitute a versatile platform for quantum state transfer, entanglement distribution and gate implementation~\cite{Bos03,Bur05,Kay06,Mar16,Sei16,Vie18,Vie19,Vie20,Kat23}. Spin-$\tfrac{1}{2}$ chains with Ising- or $XXZ$-type interactions are realized in several architectures, including trapped ions~\cite{Cir95,Mue08,Bal16,Web18}, superconducting qubits~\cite{Hou12,Zhu17}, quantum dots~\cite{Los98,Baa17} and cold atoms in optical lattices~\cite{Lew07}. These systems naturally support local control fields, which can be exploited to engineer effective few-qubit gates by driving selected spins~\cite{Heu10,Mic10}. In this context, short spin chains of two and three qubits with nearest-neighbor couplings described by a driven Hamiltonian are directly compatible with current experimental platforms and suitable for realizing two- or three-qubit controlled gates such as the Barenco, CNOT and Toffoli operations.

Building on this, driven spin chains emerge as a natural and flexible platform for multi-qubit conditional operations. In particular, the Barenco gate corresponds to a controlled unitary acting on a single target qubit, where an arbitrary $SU(2)$ transformation depends on the state of $N-1$ control qubits~\cite{Bar95}. Unlike standard approaches based on decompositions into many elementary two-qubit gates, which generally incur significant overhead for conditional $SU(2)$ operations, our protocol implements the Barenco gate directly at the Hamiltonian level, thereby avoiding explicit gate decompositions and significantly reducing the effective circuit depth.

In this work, we show how to implement Barenco gates using driven spin chains composed of two and three qubits. For two qubits, we consider an Ising interaction between the qubits together with a time-dependent transverse field on the second spin. For three qubits, we combine an $XXZ$ interaction between the first two spins with a driven Ising coupling between the second and third spins. In both cases, we identify relevant subspaces, transform to convenient rotating frames, and apply a rotating-wave approximation to derive effective Hamiltonians that directly implement Barenco gates. We obtain explicit parameter constraints, including the relation between the exchange coupling $J$ and the Ising coupling $\beta$ for the three-qubit case.

The paper is organized as follows. In Sec.~\ref{sec:background}, we review the definition of the Barenco gate, introduce the spin-chain Hamiltonians used in our protocol, and define the operator fidelity used to quantify the quality of the implementation. In Sec.~\ref{sec:results}, we present the main analytical results: we first construct the two-qubit Barenco gate (including CNOT) and then extend the protocol to three qubits (including Toffoli). We also discuss parameter constraints and the structure of the effective dynamics under disorder. Section~\ref{sec:conclusions} summarizes our findings and outlines possible extensions.

\section{Protocol background}\label{sec:background}

In this section we set up the underlying mathematical framework of our protocol. We begin by revisiting the Barenco gate and its matrix representation. Next, we introduce the driven Ising spin-chain Hamiltonians that implement the gate dynamics, and define the dimensionless parameters used to express the couplings and drive. Finally, we specify the operator fidelity employed as our figure of merit to quantify how closely the realized evolution matches the target Barenco operation.

\subsection{Barenco gate}

Quantum logic circuits are composed of quantum gates that act on registers of an arbitrary number $N$ of qubits~\cite{Nie00}. In 1995, Barenco~\cite{Bar95} introduced a family of multi-qubit controlled gates in which the state of the $N^{\text{th}}$ qubit is transformed conditionally on the joint state of the first $N-1$ qubits. In a suitable computational basis, the matrix form of the Barenco gate is
\begin{equation}
V_N(\varphi,\omega,\phi)=
\left(
\begin{array}{ccccc}
1 & 0 & \dotsb & 0 & 0\\
0 & 1 & \dotsb & 0 & 0\\
\vdots & \vdots & \ddots & \vdots & \vdots\\
0 & 0 & 0 &
e^{i\omega} \cos\!\left(\dfrac{\varphi}{2}\right) &
-i e^{i(\omega-\phi)} \sin\!\left(\dfrac{\varphi}{2}\right)\\
0 & 0 & 0 &
-i e^{i(\omega+\phi)} \sin\!\left(\dfrac{\varphi}{2}\right) &
e^{i\omega} \cos\!\left(\dfrac{\varphi}{2}\right)
\end{array}
\right),
\label{Bar}
\end{equation}
where $\varphi$, $\omega$ and $\phi$ are real parameters. If the first $N-1$ qubits are in the state $\ket{\downarrow\downarrow\ldots\downarrow}$, the $N^{\text{th}}$ (target) qubit undergoes a rotation on the Bloch sphere whose amplitudes and phases are determined by $(\varphi,\omega,\phi)$. For suitable choices of these parameters, one recovers, for instance, the 2-qubit CNOT gate and the 3-qubit Toffoli gate~\cite{Bar95,Bar95a,Tof80}. In particular, the Toffoli gate is obtained for $\varphi=\pi$, $\omega=\pi/2$ and $\phi=0$.

\subsection{Driving Hamiltonian}

The elementary building block of our protocol is an Ising spin chain with a local time-dependent transverse drive on the last spin. For an $N$-qubit chain we consider
\begin{equation}
H^{\text{drive}}_N
= \beta\, \sigma_{N-1}^z \sigma_N^z
+ \alpha\, \sigma_N^z
+ \frac{\varphi A}{\pi}
\cos\!\left(\frac{2(\beta-\alpha)t}{\hbar}-\phi\right)\sigma_N^x,
\label{Hdrive}
\end{equation}
where $\sigma^x_j$ and $\sigma^z_j$ are Pauli operators acting on qubit $j$, $\alpha$ and $\beta$ are static coupling parameters, $A$ is an energy scale, and the last term describes a transverse oscillating magnetic field applied to the $N^{\text{th}}$ spin. The drive frequency is chosen resonant with the energy difference between relevant eigenstates of the static part.

\subsection{Dimensionless parameters}

It is convenient to introduce the dimensionless time
\begin{equation}
\tau = \frac{A t}{\hbar},
\label{eq:tau_def}
\end{equation}
so that the intended gate time will simply be $\tau=\pi$. We also set
\begin{equation}
\alpha = mA,\qquad \beta = kA,
\qquad m,k\in\mathds{Z},
\label{eq:alphabeta_dimless}
\end{equation}
and write $J=\gamma A$ when needed. Throughout we assume
\begin{equation}
0\leq \varphi \leq \pi,\qquad
\alpha > \varphi,\qquad
|k-m|>1,
\label{eq:RWA_conditions}
\end{equation}
so that the rotating-wave approximation (RWA) used~\cite{Wu07} below is justified.

In terms of $\tau$, the time-dependent drive is conveniently written as
\begin{equation}
\Omega(\tau) = \frac{\varphi A}{\pi}
\cos\!\left(2(k-m)\tau-\phi\right),
\label{OmegaDef}
\end{equation}
and we will express the N-qubit Hamiltonians as a static part $H_{NQ}^{(0)}$ plus a drive,
\begin{equation}
H(t) = H_{NQ}^{(0)} + \Omega(\tau)\,\hat{V},
\end{equation}
with a simple operator $\hat{V}$ specifying which spin is driven.

\subsection{Fidelity of quantum operators}

To quantify how close a physically implemented unitary $U$ is to a target unitary $U_0$ acting on a Hilbert space of dimension $D = 2^N$ on pure input states, a natural metric is the average fidelity proposed in Ref.~\cite{Ped07} between the output states $U|\psi\rangle$ and $U_0|\psi\rangle$, averaged uniformly over all pure input states $|\psi\rangle$
\begin{equation}
F = \frac{\left| \mathrm{Tr}\!\left(M\right) \right|^2+\mathrm{Tr}\!\left(MM^\dagger\right)}{D(D+1)},
\label{Fid}
\end{equation}
where $M=U_0^\dagger U$. As $U$ and $U_0$ are unitary operators, we get $\mathrm{Tr}\!\left(MM^\dagger\right)=D$, and we can rewrite the Fidelity as $F = (Df+1)/(D+1)$ where $f = |\mathrm{Tr}(M)|^2/D^2$ and $0 \leq f \leq 1$ \cite{Cab10}. So, by construction, $1/(D+1) \leq F \leq 1$. The value $F = 1$ is achieved if and only if $U$ and $U_0$ differ at most by a global phase, whereas $F = 1/(D+1)$ corresponds to the case in which $U$ and $U_0$ are maximally distinguishable on average over pure input states. In what follows, $U_0$ will be the ideal Barenco operator $V_N(\varphi,\omega,\phi)$ and $U$ the time-evolution operator generated by the spin-chain Hamiltonian at the chosen gate time. Throughout, we neglect overall global phases common to all computational-basis states, as they have no physical effect.

%%%%%%%%%%%%%%%%%%%%%%%%%%%%%%%%%%%%%%%%%%%%%%%%%%%%%%%%%%%%
\section{Results}\label{sec:results}

In this section we present our main analytical results. We first construct the two-qubit Barenco gate from a driven Ising chain and then extend the protocol to three qubits using an $XXZ$ interaction and the same driven structure on the last spin. 

\subsection{Implementation of the two-qubit Barenco gate}\label{sec:twoqubit}

We consider a two–spin chain with a $ZZ$ coupling between a control and a target qubit, while only the target is driven transversely. In a suitable rotating frame the dynamics splits into two independent $2\times2$ blocks set by the control state; one block is idle (up to a global phase) and the other undergoes a driven Rabi rotation. By matching the drive amplitude, phase, and duration, this realizes the Barenco gate $V_2(\varphi,\omega,\phi)$ at a fixed gate time, which we benchmark via operator fidelity against the ideal form. Our calculations are shown as follows.

\subsubsection{Model Hamiltonian}

For two qubits, we consider the Hamiltonian
\begin{align}
H
&= \frac{\omega A}{2\pi}\,\sigma_1^z + H^{\text{drive}}_2 \nonumber\\
&= \frac{\omega A}{2\pi}\,\sigma_1^z
   + \beta\, \sigma_1^z \sigma_2^z
   + \alpha\, \sigma_2^z
   + \frac{\varphi A}{\pi}
     \cos\!\left(\frac{2(\beta-\alpha)t}{\hbar}-\phi\right)\sigma_2^x.
\label{Hcnot}
\end{align}
The first term is a local $z$ field on qubit~1, the next two describe a static Ising coupling and $z$ field on qubit~2, and the last term is a transverse drive on the second qubit.

Using the dimensionless time $\tau$ introduced in Eq.~\eqref{eq:tau_def}, the drive reads $\Omega(\tau)$ as in Eq.~\eqref{OmegaDef}. In the computational basis
\(
\{\ket{\downarrow\downarrow},
  \ket{\downarrow\uparrow},
  \ket{\uparrow\downarrow},
  \ket{\uparrow\uparrow}\}
\),
it is convenient to write
\begin{equation}
H_{\text{2Q}}(\tau)
= H_{\text{2Q}}^{(0)} + \Omega(\tau)\,(\sigma_x \oplus \sigma_x),
\label{H2Q_split}
\end{equation}
where the static part is
\begin{equation}
H_{\text{2Q}}^{(0)} =
\mathrm{diag}\!\left(
\alpha+\beta+\frac{\omega A}{2\pi},
-\alpha-\beta+\frac{\omega A}{2\pi},
\alpha-\beta-\frac{\omega A}{\pi},
-\alpha+\beta-\frac{\omega A}{\pi}
\right),
\label{H2Q0}
\end{equation}
and
\begin{equation}
\sigma_x \oplus \sigma_x =
\begin{pmatrix}
0 & 1 & 0 & 0\\
1 & 0 & 0 & 0\\
0 & 0 & 0 & 1\\
0 & 0 & 1 & 0
\end{pmatrix}.
\label{sigmaxoplus}
\end{equation}

\subsubsection{Rotating frame and block structure}

To remove trivial single-qubit phases due to the static $z$ fields, we perform a time-dependent unitary transformation
\begin{equation}
H_{\text{2Q}}(\tau)
= \hat{U}(\tau)\,\mathcal{H}_{\text{2Q}}(\tau)\,\hat{U}^\dagger(\tau)
 - i\hbar\,\frac{d\hat{U}(\tau)}{dt}\,\hat{U}^\dagger(\tau),
\label{UT}
\end{equation}
with
\begin{equation}
\hat{U}(\tau)
= e^{\frac{i}{\hbar}\left(\frac{\omega A}{2\pi}\right)t}
\begin{pmatrix}
1 & 0 & 0 & 0\\
0 & 1 & 0 & 0\\
0 & 0 & e^{-i\frac{\omega A t}{2\pi\hbar}} & 0\\
0 & 0 & 0 & e^{-i\frac{\omega A t}{2\pi\hbar}}
\end{pmatrix}.
\label{U0}
\end{equation}
In this rotating frame, the Hamiltonian takes a block-diagonal form
\begin{equation}
H_{\text{2Q}}(\tau)
= H_{\text{upper}}(\tau)\oplus H_{\text{lower}}(\tau),
\end{equation}
where
\begin{equation}
H_{\text{upper}}(\tau)=
\begin{pmatrix}
\alpha+\beta &
\Omega(\tau)\\[1mm]
\Omega(\tau) &
-\alpha-\beta
\end{pmatrix},
~
H_{\text{lower}}(\tau)=
\begin{pmatrix}
\alpha-\beta-\dfrac{\omega A}{\pi} &
\Omega(\tau)\\[1mm]
\Omega(\tau) &
-\alpha+\beta-\dfrac{\omega A}{\pi}
\end{pmatrix},
\label{Hupdownblocks}
\end{equation}
and we recall that
\begin{equation}
\Omega(\tau)=\frac{\varphi A}{\pi}
\cos\!\left(2(k-m)\tau-\phi\right).
\end{equation}
The block $H_{\text{upper}}(\tau)$ acts on the subspace
\begin{equation}
\mathcal{H}_{\text{upper}} =
\operatorname{span}\{\ket{\uparrow\uparrow},\ket{\uparrow\downarrow}\},
\end{equation}
while $H_{\text{lower}}(\tau)$ acts on
\begin{equation}
\mathcal{H}_{\text{lower}} =
\operatorname{span}\{\ket{\downarrow\uparrow},\ket{\downarrow\downarrow}\}.
\end{equation}
This separation isolates the computationally relevant subspace and makes the effective dynamics exactly solvable.

\subsubsection{Effective dynamics: upper block}

We first consider $H_{\text{upper}}(\tau)$. Writing $\cos\theta=\tfrac{1}{2}(e^{i\theta}+e^{-i\theta})$ and introducing
\begin{equation}
\theta(\tau) = 2(k-m)\tau-\phi,
\end{equation}
we apply a further unitary transformation
\begin{equation}
\hat{U}_1(t)=
\begin{pmatrix}
e^{i (\alpha+\beta)t/\hbar} & 0\\[1mm]
0 & e^{-i (\alpha+\beta)t/\hbar}
\end{pmatrix}
\end{equation}
to eliminate the static diagonal terms in $H_{\text{upper}}(\tau)$. In this interaction picture, the transformed Hamiltonian contains rapidly oscillating terms proportional to $e^{\pm i(2(\beta-\alpha)\pm 2(\alpha+\beta))t/\hbar}$, which average out when the RWA conditions in Eq.~\eqref{eq:RWA_conditions} are satisfied. Keeping only the slowly varying contributions~\cite{Wu07,Kri19}, the effective Hamiltonian for the upper block reduces to a multiple of the identity, which generates only a global phase. We therefore take
\begin{equation}
\hat{H}_{\text{upper}}^{\text{(eff)}}=
\begin{pmatrix}
0 & 0\\
0 & 0
\end{pmatrix},
\label{HupEff}
\end{equation}
so that
\begin{equation}
U_{\text{upper}}(t)
= e^{-i \hat{H}_{\text{upper}}^{\text{(eff)}} t/\hbar}
=
\begin{pmatrix}
1 & 0\\
0 & 1
\end{pmatrix}.
\end{equation}
Thus the subspace with the first qubit in $\ket{\uparrow}$ is essentially frozen in this frame.

\subsubsection{Effective dynamics: lower block}

The nontrivial dynamics arises in the lower block $H_{\text{lower}}(\tau)$. We apply the unitary
\begin{equation}
\hat{U}_3(t)=
\begin{pmatrix}
e^{i (\alpha-\beta)t/\hbar} & 0  \\
0 & e^{-i (\alpha-\beta)t/\hbar}
\end{pmatrix}
e^{-i\frac{\omega A}{\pi\hbar}t}
\end{equation}
to remove the diagonal terms in \eqref{Hupdownblocks}. In the corresponding interaction picture we obtain
\begin{equation}
\hat{H}_{\text{lower}}^{\text{(trans)}}(t)=
\begin{pmatrix}
0 & \displaystyle \frac{\varphi A}{2\pi} \bigl(e^{i\frac{4(\beta-\alpha) t}{\hbar}}e^{-i\phi}+e^{i\phi}\bigr)\\[2mm]
\displaystyle \frac{\varphi A}{2\pi} \bigl(e^{-i\frac{4(\beta-\alpha) t}{\hbar}}e^{i\phi}+e^{-i\phi}\bigr) & 0
\end{pmatrix}.
\label{Hh3}
\end{equation}
Neglecting the rapidly oscillating terms $e^{\pm i 4(\beta-\alpha)t/\hbar}$ under the RWA, we obtain the effective Hamiltonian
\begin{equation}
\hat{H}_{\text{lower}}^{\text{(eff)}}=\frac{\varphi A}{2\pi}
\begin{pmatrix}
0 & e^{-i\phi}\\
e^{i\phi} & 0
\end{pmatrix}
= \frac{\varphi A}{2\pi}\,
\bigl(\cos\phi\,\sigma_x+\sin\phi\,\sigma_y\bigr),
\label{Hh3080}
\end{equation}
which generates a rotation around an axis in the equatorial plane of the Bloch sphere. The corresponding time-evolution operator is
\begin{equation}
e^{-i \hat{H}_{\text{lower}}^{\text{(eff)}} t/\hbar}=
\begin{pmatrix}
\cos\!\left(\dfrac{A\varphi t}{2\pi\hbar}\right) &
-i e^{-i\phi}\sin\!\left(\dfrac{A\varphi t}{2\pi\hbar}\right)\\[2mm]
-i e^{i\phi}\sin\!\left(\dfrac{A\varphi t}{2\pi\hbar}\right) &
\cos\!\left(\dfrac{A\varphi t}{2\pi\hbar}\right)
\end{pmatrix}.
\label{part1}
\end{equation}
This is a generic $SU(2)$ rotation around the equatorial axis
$\boldsymbol{n} = (\cos\phi,\sin\phi,0)$
with angular frequency $\varphi A/(2\pi\hbar)$.

Transforming back to the original rotating frame, the evolution operator in the lower block reads
\begin{equation}
U_{\text{lower}}(t)
\!=\! e^{i \frac{\omega At}{\pi\hbar}}
\!
\begin{pmatrix}
e^{i\frac{(\alpha\!-\!\beta)t}{\hbar}} & 0  \\
0 & e^{-i\frac{(\alpha\!-\!\beta)t}{\hbar}}
\end{pmatrix}\!
\begin{pmatrix}
\!\cos\!\left(\dfrac{A\varphi t}{2\pi\hbar}\right) &
\!-i e^{-i\phi}\sin\!\left(\dfrac{A\varphi t}{2\pi\hbar}\right)\\[2mm]
\!-i e^{i\phi}\sin\!\left(\dfrac{A\varphi t}{2\pi\hbar}\right) &
\!\cos\!\left(\dfrac{A\varphi t}{2\pi\hbar}\right)
\end{pmatrix}.
\label{BA1}
\end{equation}

\subsubsection{Gate time and Barenco operator}

We now choose the gate time
\begin{equation}
\tau_{\text{g}} = \frac{A t_{\text{g}}}{\hbar} = \pi,
\qquad\Rightarrow\qquad
t_{\text{g}} = \frac{\pi\hbar}{A}.
\label{taudef}
\end{equation}
Using $\alpha=mA$ and $\beta=kA$ with $m,k\in\mathds{Z}$, we have
\begin{equation}
e^{i(\alpha-\beta)t_{\text{g}}/\hbar} = e^{i(m-k)\pi} = (-1)^{m-k},
\qquad
e^{i\omega A t_{\text{g}}/(\pi\hbar)} = e^{i\omega}.
\end{equation}
Up to a global phase $(-1)^{m-k}$, the action of $U_{\text{lower}}(t_{\text{g}})$ on the lower subspace becomes
\begin{equation}
U_{\text{lower}}(t_{\text{g}}) =
\begin{pmatrix}
e^{i\omega}\cos\left(\dfrac{\varphi}{2}\right) &
- i e^{i(\omega+\phi)} \sin\left(\dfrac{\varphi}{2}\right)\\[2mm]
- i e^{i(\omega-\phi)} \sin\left(\dfrac{\varphi}{2}\right) &
e^{i\omega}\cos\left(\dfrac{\varphi}{2}\right)
\end{pmatrix}.
\end{equation}
Since $U_{\text{upper}}(t_{\text{g}})=\mathbb{1}_2$, the full two-qubit evolution operator in the computational basis coincides with the two-qubit Barenco gate $V_2(\varphi,\omega,\phi)$:
\begin{equation}
U(t_{\text{g}})\hspace{-.1cm}=
\left(
\begin{array}{cccc}
1 & 0 & 0 & 0\\
0 & 1 & 0 & 0\\
0 & 0 & e^{i\omega} \cos\left(\dfrac{\varphi}{2}\right)
      & -i e^{i(\omega+\phi)} \sin\left(\dfrac{\varphi}{2}\right)\\[1mm]
0 & 0 & -i e^{i(\omega-\phi)} \sin\left(\dfrac{\varphi}{2}\right)
      & e^{i\omega} \cos\left(\dfrac{\varphi}{2}\right)
\end{array}
\right).
\label{matrij}
\end{equation}

\subsubsection{Examples and additional remarks}

For example, choosing
\begin{equation}
\omega=\frac{\pi}{2},\qquad \varphi=\pi,\qquad \phi=0,
\end{equation}
we obtain the CNOT gate with the first qubit as control and the second as target. More generally, Eq.~\eqref{Hh3080} shows that the effective Hamiltonian in the lower block implements a rotation around an axis in the equatorial plane with angle $\varphi$; the phase $\phi$ fixes the rotation axis in the $x$--$y$ plane, while $\omega$ sets the overall $z$-phase of the target qubit. Thus the protocol realizes a controlled-$SU(2)$ rotation with three tunable parameters, precisely matching the Barenco family. %%% Discutir artigo do Zinner.

%%%%%%%%%%%%%%%%%%%%%%%%%%%%%%%%%%%%%%%%%%%%%%%%%%%%%%%%%%%%
\subsection{Implementation of the three-qubit Barenco gate}\label{sec:threequbit}

We now extend the previous construction to three qubits. The first two qubits are coupled through an $XXZ$ interaction and subject to identical local $z$ fields, while the second and third qubits are coupled via the driven Ising Hamiltonian of Eq.~\eqref{Hdrive}.

\subsubsection{XXZ gate between the first two qubits}

We consider the $XXZ$ Hamiltonian
\begin{align}
H_{XXZ}
&= \frac{J}{2}\left[\sigma_{1}^x\sigma_{2}^x
                 + \sigma_{1}^y\sigma_{2}^y
                 + \left(1-\frac{\omega A}{2\pi J}\right)\sigma_{1}^z\sigma_{2}^z\right]
   + \frac{\omega A}{4\pi}\left(\sigma_{1}^z+\sigma_{2}^z\right),
\label{H3}
\end{align}
where $J$ is an exchange coupling. If the evolution time satisfies
\begin{equation}
\frac{Jt}{\hbar} = n\pi, \qquad n\in\mathds{N},
\end{equation}
the time evolution generated by $H_{XXZ}$ implements a controlled-phase gate
\begin{equation}
U(\tau) = G(\omega) =
\left(
\begin{array}{cccc}
1 & 0 & 0 & 0 \\
0 & 1 & 0 & 0 \\
0 & 0 & 1 & 0 \\
0 & 0 & 0 & e^{i\omega}
\end{array}
\right),
\end{equation}
in the basis
$\{\ket{\downarrow\downarrow},
  \ket{\downarrow\uparrow},
  \ket{\uparrow\downarrow},
  \ket{\uparrow\uparrow}\}$ of the first two qubits.

\subsubsection{Three-qubit Hamiltonian and block structure}

Including the driven interaction with the third qubit, the full Hamiltonian reads
\begin{equation}
H = H_{XXZ} + H^{\text{drive}}_3,
\end{equation}
i.e.
\begin{align}
H
&= \frac{J}{2}\left[\sigma_{1}^x\sigma_{2}^x
                 + \sigma_{1}^y\sigma_{2}^y
                 + \left(1-\frac{\omega A}{2\pi J}\right)\sigma_{1}^z\sigma_{2}^z\right]
   + \frac{\omega A}{4\pi}\left(\sigma_{1}^z+\sigma_{2}^z\right) \nonumber\\
&\quad + \beta\, \sigma_{2}^z \sigma_3^z
       + \alpha\, \sigma_3^z
       + \frac{\varphi A}{\pi}
         \cos\!\left(\frac{2(\beta-\alpha)t}{\hbar}-\phi\right)\sigma_3^x.
\label{Ha00}
\end{align}
In terms of $\tau$, the drive is again $\Omega(\tau)$, and we can write
\begin{equation}
H(t) = H_{\text{3Q}}^{(0)} + \Omega(\tau)\,
\bigl(\mathbb{1}\otimes\mathbb{1}\otimes\sigma_x\bigr).
\label{H3Q_split}
\end{equation}

In the three-qubit computational basis
\begin{align}
\Bigl\{\ket{\downarrow\downarrow\downarrow},
      \ket{\downarrow\downarrow\uparrow},
      \ket{\downarrow\uparrow\downarrow},
      \ket{\downarrow\uparrow\uparrow}, \nonumber\\
      \ket{\uparrow\downarrow\downarrow},
      \ket{\uparrow\downarrow\uparrow},
      \ket{\uparrow\uparrow\downarrow},
      \ket{\uparrow\uparrow\uparrow}\Bigr\},
\end{align}
the static part $H_{\text{3Q}}^{(0)}$ is an $8\times 8$ matrix. After separating global phases using a suitable unitary transformation (slight similar to Eq.~\eqref{UT}), the Hamiltonian exhibits a block structure consisting of three diagonal blocks: a $2\times 2$ block associated with the subspace $\{\ket{\uparrow\downarrow\downarrow},\ket{\uparrow\downarrow\uparrow}\}$, a $2\times 2$ block associated with $\{\ket{\downarrow\downarrow\downarrow},\ket{\downarrow\downarrow\uparrow}\}$, and a $4\times 4$ block $H_2(t)$ associated with the subspace
\begin{equation}
\mathcal{H}_2 = \operatorname{span}\Bigl\{\ket{\downarrow\downarrow\uparrow},
\ket{\downarrow\uparrow\uparrow},
\ket{\uparrow\downarrow\uparrow},
\ket{\uparrow\uparrow\uparrow}\Bigr\}.
\end{equation}
The first two blocks are equivalent (up to phases) to the upper and lower $2\times 2$ blocks in the two-qubit analysis and do not contribute nontrivially to the desired three-qubit gate. The essential dynamics is encoded in $H_2(t)$, which we now analyze.

\subsubsection{Four-dimensional block and effective dynamics}

Within the subspace $\mathcal{H}_2$, the Hamiltonian block can be written as
\begin{equation}
H_{2}(t)=
\left(
\begin{array}{cccc}
\alpha-\beta-J &
\Omega(\tau) &
J & 0\\[2mm]
\Omega(\tau) &
-\alpha+\beta-J &
0 & J\\[2mm]
J & 0 &
\alpha+\beta-J &
\Omega(\tau)\\[2mm]
0 & J &
\Omega(\tau) &
-\alpha-\beta-J
\end{array}
\right).
\label{H2Block}
\end{equation}
If we initially neglect the drive, the static part $H_{\text{2 input}}$ is
\begin{equation}
H_{\text{2 input}}=
\begin{pmatrix}
\alpha-\beta-J & 0 & J & 0\\
0 & -\alpha+\beta-J & 0 & J\\
J & 0 & \alpha+\beta-J & 0\\
0 & J & 0 & -\alpha-\beta-J
\end{pmatrix}.
\label{H2input}
\end{equation}
Diagonalizing $H_{\text{2 input}}$ yields eigenvalues
\begin{equation}
E_{\pm}^{(\pm)} = \pm\left(\alpha\pm\sqrt{J^2+\beta^2}\right).
\end{equation}
The corresponding eigenvectors form a basis in which the static part is diagonal. In that basis, the transverse drive couples pairs of eigenstates separated by an energy difference controlled by $\sqrt{J^2+\beta^2}$.

Proceeding as in the two-qubit case, we move to an interaction picture defined by the eigenbasis of $H_{\text{2 input}}$ and apply a rotating-wave approximation that retains only near-resonant couplings. This results in an effective diagonal evolution operator
\small{
\begin{equation}
\hat{U}_2(t)\!=\!
\begin{pmatrix}
\!e^{\!-\frac{i}{\hbar}\left(\alpha\!+\!\sqrt{J^2\!+\!\beta^2}\right)t} & \!0 & \!0 & \!0 \\
\!0 & \!e^{\!-\frac{i}{\hbar}\left(\alpha\!-\!\sqrt{J^2\!+\!\beta^2}\right)t} & \!0 & \!0 \\
\!0 & \!0 & e^{\!\frac{i}{\hbar}\left(\alpha\!-\!\sqrt{J^2\!+\!\beta^2}\right)t} & \!0 \\
\!0 & \!0 & \!0 & \!e^{\!\frac{i}{\hbar}\left(\alpha\!+\!\sqrt{J^2\!+\!\beta^2}\right)t}
\end{pmatrix}\!
e^{\!-iJt/\hbar},
\label{U2block}
\end{equation}}
acting on $\mathcal{H}_2$. At the gate time $t=t_{\text{g}}=\pi\hbar/A$, we impose the condition that the phases corresponding to $\pm\sqrt{J^2+\beta^2}$ coincide modulo $2\pi$:
\begin{align}
e^{\frac{i}{\hbar}\sqrt{J^2+\beta^2}t_{\text{g}}}
&= e^{-\frac{i}{\hbar}\sqrt{J^2+\beta^2}t_{\text{g}}}, \nonumber\\
e^{2\frac{i}{\hbar}\sqrt{J^2+\beta^2}t_{\text{g}}}
&= 1,
\end{align}
which implies
\begin{equation}
2\frac{1}{\hbar}\sqrt{J^2+\beta^2}t_{\text{g}} = 2\pi l,
\qquad l\in\mathds{Z},
\end{equation}
and hence
\begin{equation}
\sqrt{J^2+\beta^2} = lA
\quad\Longrightarrow\quad
\left(\frac{J}{A}\right)^2 + k^2 = l^2,
\end{equation}
where we used $\beta=kA$. Thus the dimensionless couplings satisfy a simple relation. Equivalently,
\begin{equation}
J = \pm\sqrt{l^2-k^2}\,A,
\label{Jcondition}
\end{equation}
with integers $l>|k|$. Equation~\eqref{Jcondition} is an additional constraint on $J$ that ensures the desired phase structure in the block $\mathcal{H}_2$ at the gate time.

\subsubsection{Three-qubit Barenco and Toffoli gate}

Combining the block evolution in $\mathcal{H}_2$ with the trivial evolution in the remaining subspaces, and accounting for the driven dynamics of the third spin as in the two-qubit case, the full three-qubit time-evolution operator at $t=t_{\text{g}}$ can be written as
\begin{equation}
U(t_{\text{g}})=
\left(
\begin{array}{cccccccc}
1 & 0 & 0 & 0 & 0 & 0 & 0 & 0\\
0 & 1 & 0 & 0 & 0 & 0 & 0 & 0\\
0 & 0 & 1 & 0 & 0 & 0 & 0 & 0\\
0 & 0 & 0 & 1 & 0 & 0 & 0 & 0\\
0 & 0 & 0 & 0 & 1 & 0 & 0 & 0\\
0 & 0 & 0 & 0 & 0 & 1 & 0 & 0\\
0 & 0 & 0 & 0 & 0 & 0 &
e^{i\omega} \cos\!\left(\dfrac{\varphi}{2}\right) &
- i e^{i(\omega+\phi)} \sin\!\left(\dfrac{\varphi}{2}\right)\\[1mm]
0 & 0 & 0 & 0 & 0 & 0 &
- i e^{i(\omega-\phi)} \sin\!\left(\dfrac{\varphi}{2}\right) &
e^{i\omega} \cos\!\left(\dfrac{\varphi}{2}\right)
\end{array}
\right),
\label{matr}
\end{equation}
up to an overall global phase. This is precisely the three-qubit Barenco gate $V_3(\varphi,\omega,\phi)$ with the first two qubits acting as controls and the third as target.

In particular, choosing
\begin{equation}
\omega=\frac{\pi}{2},\qquad \varphi=\pi,\qquad \phi=0,
\end{equation}
we obtain the three-qubit Toffoli gate. Equation~\eqref{Jcondition} shows that for any integer $l>|k|$ one can choose $J=\pm\sqrt{l^2-k^2}\,A$ to satisfy the resonance condition, providing a discrete family of parameter sets that implement the same logical gate.

%%%%%%%%%%%%%%%%%%%%%%%%%%%%%%%%%%%%%%%%%%%%%%%%%%%%%%%%%%%%
\subsection{Fidelity and numerical results}

To assess the performance and robustness of the protocol, we compute the operator fidelity~\eqref{Fid} between the ideal Barenco gate $V_N(\varphi,\omega,\phi)$ and the exact time-evolution operator $U(t)$ generated by the full Hamiltonian of each case (two qubits: Eq.~\eqref{Hcnot}; three qubits: Eq.~\eqref{Ha00}). We sample a range of parameter values $0\le \varphi \le \pi$, $0\le \omega < 2\pi$, $0\le \phi < 2\pi$, with incremental steps of $10\%$ of each interval, yielding $11\times 10\times 10=1100$ parameter triples. For each triple we propagate $U(t)$ and compute $F(t)$, then average over all triples to obtain $\braket{F(t)}$.

\begin{figure}[!ht]
\begin{center}
\includegraphics[width=0.6\linewidth]{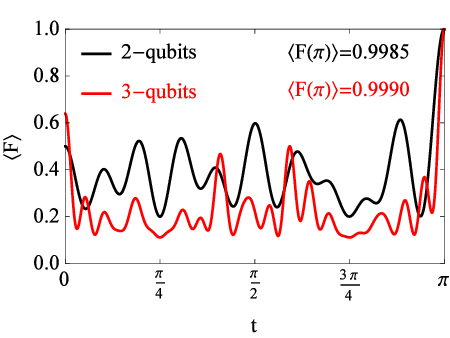}
\caption{
Average fidelity $\braket{F(t)}$ for the two- and three-qubit spin-chain implementations of the Barenco operators $V_2(\varphi,\omega,\phi)$ and $V_3(\varphi,\omega,\phi)$ as a function of time. We average $1100$ realizations by varying $0\le \varphi \le \pi$, $0\le \omega < 2\pi$, and $0\le \phi < 2\pi$ with steps of $10\%$ of each range. Parameters: $\alpha=2A$ for both cases, and $J=\sqrt{l^2-k^2}\,A$ with $l=17$ and $k=8$ for the three-qubit case. The curves reach $\braket{F(\pi)}>0.998$, illustrating efficiency and robustness of the protocol.}
\label{Fig1}
\end{center}
\end{figure}

Figure~\ref{Fig1} displays $\braket{F(t)}$ for both implementations on the same time axis (units of $\pi\hbar/A$, so $t=\pi\hbar/A$ corresponds to the gate time $t_{\text{g}}$). For all data we set $\alpha=2A$. In the three-qubit case we additionally impose
\begin{equation}
\frac{J}{A}=\sqrt{l^2-k^2},
\end{equation}
with $l=17$ and $k=8$, consistent with the resonance condition~\eqref{Jcondition}. The two-qubit Hamiltonian does not involve $J$.

Although Fig.~\ref{Fig1} displays an average over parameter sets, simulations for individual choices of $(\varphi,\omega,\phi)$ show essentially the same temporal behavior, indicating that the fidelity profile is largely independent of the precise values of these parameters within the explored range. This highlights both the tunability and robustness of the protocol.

\subsubsection{Disorder analysis}

\begin{figure}[!ht]
\begin{center}
\includegraphics[width=\linewidth]{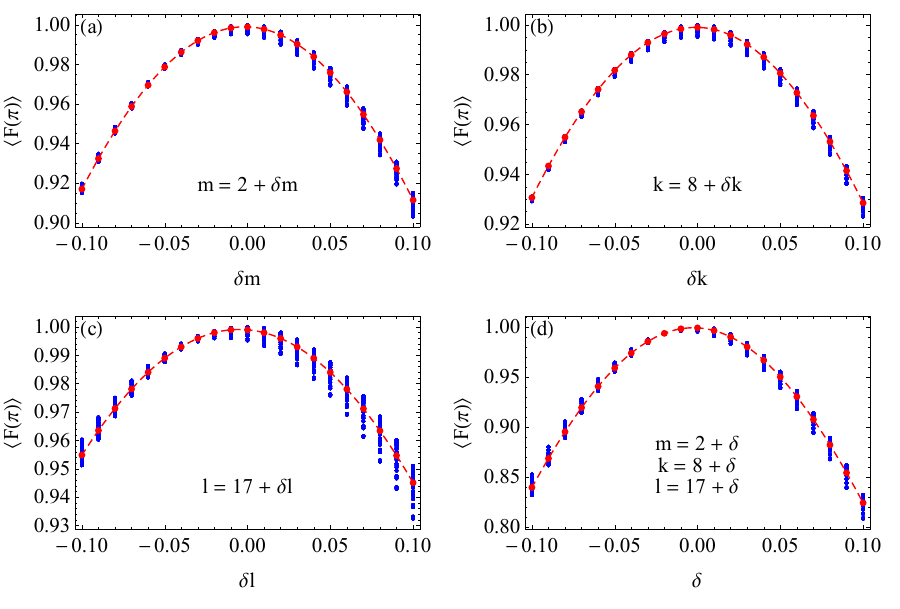}
\caption{
Disorder analysis with variations $m\!\to\! m+\delta m$, $k\!\to\! k+\delta k$, and $l\!\to\! l+\delta l$ about $(m,k,l)=(2,8,17)$. For each $\delta$ value (from $-0.10$ to $0.10$ in steps of $0.01$), we average $1100$ realizations (blue points) over $(\varphi,\omega,\phi)$ as in Fig.~\ref{Fig1}. Red points and the dashed line indicate per-$\delta$ averages. Panels: (a)~$\delta m$ only; (b)~$\delta k$ only; (c)~$\delta l$ only; (d)~joint variation $\delta m=\delta k=\delta l=\delta$. Even with disorder, average fidelities remain above $0.99$ for $|\delta|<0.01$ across all panels, indicating strong robustness.}
\label{Fig2}
\end{center}
\end{figure}

The analytical construction of the three-qubit Barenco gate relies on integer triplets $(m,k,l)$ that set the parameters $\alpha=mA$, $\beta=kA$, and $J=\sqrt{l^2-k^2}\,A$. To probe robustness against fabrication and control imperfections, we introduce small deviations
\begin{align}
\alpha(m+\delta m) &= (m+\delta m)A, \nonumber\\
\beta(k+\delta k)  &= (k+\delta k)A, \nonumber\\
J(l+\delta l,k+\delta k) &= \sqrt{(l+\delta l)^2-(k+\delta k)^2}\,A,
\end{align}
where $\delta m$, $\delta k$, and $\delta l$ are dimensionless perturbations. Figure~\ref{Fig2} summarizes four scenarios for the baseline $(m,k,l)=(2,8,17)$: panels (a)–(c) vary one parameter at a time; panel (d) varies all simultaneously with $\delta m=\delta k=\delta l=\delta$. For each value of $\delta\in[-0.10,0.10]$ (step $0.01$) we average $1100$ realizations over the same $(\varphi,\omega,\phi)$ grid used above.

Overall, small deviations do not significantly degrade performance across the entire parameter window explored. When varying a \emph{single} calibration constant at a time, the fidelities remain essentially unchanged within numerical precision for $|\delta j|\le 0.03$ ($j\in\{m,k,l\}$), and only begin to dip below $99\%$ for larger offsets. This trend is consistent with the structure of the effective Hamiltonians: the drive-induced rotations depend on the detuning $2(\beta-\alpha)$ and the gate-time scaling set by $A$, so modest perturbations in $\alpha$ or $\beta$ primarily rescale the local rotation rates without qualitatively altering the block structure that implements the Barenco action. In contrast, $J$ enters the three-qubit construction through the resonance constraint $J/A=\sqrt{l^2-k^2}$; a perturbation of $(l,k)$ thus modifies $J$ as
\[
\frac{\delta J}{A}\;\approx\;\frac{l\,\delta l-k\,\delta k}{\sqrt{l^2-k^2}},
\]
so that joint variations $(\delta l,\delta k)$ that add constructively (i.e., $l\,\delta l\approx k\,\delta k$) are the most detrimental. This explains the slightly stronger degradation seen when all three parameters are varied simultaneously: the effective detunings and rotation angles drift coherently rather than randomly, moving the system farther from the discrete resonance family that underpins the analytic protocol.

From a dynamical perspective, parameter shifts perturb both the diagonal energies (changing accumulated dynamical phases) and the transverse Rabi rates (changing rotation angles), thereby producing \emph{asymmetric} distortions of the trajectories in Hilbert space. This asymmetry accounts for the mild differences observed between positive and negative deviations at larger $|\delta|$, particularly near the gate time $t_{\mathrm g}=\pi\hbar/A$ where small errors in angle or phase have the largest impact on $F(t)$. Nevertheless, for realistic calibration budgets the scheme remains comfortably in a high-fidelity regime: even under \emph{simultaneous} variations with $|\delta|=0.01$ we find averaged fidelities above $99\%$, and for single-parameter drifts as large as $|\delta j|=0.03$ the fidelity profile is essentially indistinguishable from the ideal case. Practically, this robustness means that modest static inhomogeneities or slow control drift can be tolerated without active error mitigation; remaining sensitivity is dominated by coherent detuning of $2(\beta-\alpha)$ and by violations of the $J/A=\sqrt{l^2-k^2}$ resonance, which can be compensated either by fine retuning of $(m,k,l)$ or by a small, deterministic adjustment of the gate time.

\section{Conclusions}\label{sec:conclusions}

We have presented a fully analytical protocol for implementing Barenco-type multi-qubit controlled gates using short driven spin chains. For two qubits, starting from an Ising interaction and a resonant transverse drive on the second qubit, we showed how a sequence of unitary transformations and a rotating-wave approximation lead to an effective Hamiltonian that realizes the two-qubit Barenco gate $V_2(\varphi,\omega,\phi)$ at a well-defined gate time. This includes the standard CNOT gate as a particular case.

We then embedded this construction into a three-qubit $XXZ$ chain, where the first two qubits are coupled through an $XXZ$ interaction and the second and third qubits are linked by the same driven Ising scheme. By exploiting the block structure of the three-qubit Hamiltonian and carefully tuning the exchange coupling $J$ according to Eq.~\eqref{Jcondition}, we obtained an effective evolution in a four-dimensional subspace that yields the three-qubit Barenco gate $V_3(\varphi,\omega,\phi)$, including the Toffoli gate as a special case.

All derivations provide a self-contained account of the protocol. Numerical simulations based on the full Hamiltonian confirm that high fidelities can be achieved over broad parameter ranges, demonstrating the robustness of the scheme with respect to moderate variations of $(\varphi,\omega,\phi)$ and the static couplings.

Our results illustrate how physically simple spin-chain Hamiltonians with local drivings can implement nontrivial multi-qubit gates in a transparent and analytically tractable way. Possible extensions include the analysis of decoherence effects, pulse-shape optimization, and the generalization to longer chains to realize higher-order multi-controlled operations and more complex quantum circuits.

%%%%%%%%%%%%%%%%%%%%%%%%%%%%%%%%%%%%%%%%%%%%%%%%%%%%%%%%%%%%

\section*{Statements \& Declarations}

\subsection*{Funding} 

This work was supported by Conselho Nacional de Desenvolvimento Científico e Tecnológico (CNPq) through grant number 409673/2022-6. R. Vieira has received research support from Fundação de Amparo à Pesquisa e Inovação do Estado de Santa Catarina (FAPESC), Edital 25/2025.

\subsection*{Competing Interests}

The authors have no relevant financial or non-financial interests to disclose.

\subsection*{Author Contributions}

E.P.M.A. and R.V. equally contributed to this work and reviewed the manuscript.

\end{document}